\documentclass[10pt,letterpaper]{article}
\usepackage{opex3}
\usepackage{amsfonts}
\usepackage{color}
\usepackage{rotating}

\begin{document}

\title{MHz rate and efficient synchronous heralding of single photons at telecom wavelengths}

\author{Enrico Pomarico, Bruno Sanguinetti, Thiago Guerreiro, Rob Thew, and Hugo Zbinden}

\address{Group of Applied Physics, University of Geneva, CH-1211 Geneva 4, Switzerland.}

\email{enrico.pomarico@unige.ch}

\begin{abstract}
We report on the realization of a synchronous source of heralded single photons at telecom wavelengths with MHz heralding rates
and high heralding efficiency. This source is based on the generation of photon pairs at 810 and 1550\,nm via Spontaneous
Parametric Down Conversion (SPDC) in a 1\,cm periodically poled lithium niobate (PPLN) crystal pumped by a 532\,nm pulsed laser.
As high rates are fundamental for multi-photon experiments, we show that single telecom photons can be announced at 4.4\,MHz rate
with 45\% heralding efficiency. When we focus only on the optimization of the
coupling of the heralded photon, the heralding efficiency can be increased up to 80\%.
Furthermore, we experimentally observe that group velocity mismatch inside long crystals pumped in a pulsed mode affects the spectrum of the emitted photons and their fibre coupling efficiency. The length of the crystal in this source has been chosen as a trade off between high brightness and high coupling efficiency.
\end{abstract}

\ocis{ }

\section{Introduction}
In recent years, the generation of single photon Fock states has seen a significant increase in interest, not only from a fundamental perspective, but also for their key role in many quantum information applications \cite{Castelletto2008,Eisaman2011}, such as linear-optics quantum computing  \cite{Knill2001} and single-photon detectors calibration \cite{Polyakov2009}. In the context of quantum communication \cite{Gisin2007}, heralded single photon (HSP) sources could be adopted for a wide range of tasks, like device independent Quantum Key Distribution (QKD) \cite{Gisin2010} or distribution of entangled photons at a distance via quantum repeaters \cite{Sangouard2007, Sangouard2011a}.

The idea of a HSP source was introduced for the first time in 1986, when Hong and Mandel~\cite{Hong86} showed that a localized one-photon Fock state can be achieved by exploiting the temporal correlation of the photon pairs generated by Spontaneous Parametric Down Conversion (SPDC): one photon can be heralded by the detection of its twin.
Due to the probabilistic nature of SPDC, single photons cannot be generated on demand. However, upon detection of one of the twin photons
at a heralding rate $R^H$, one can define the heralding efficiency $\eta^{H}$, i.e. the probability of finding a photon in a single mode
fibre given the trigger signal. Thus far, significant theoretical and experimental efforts have been made to efficiently couple into fibre
single photons generated via SPDC in bulk \cite{Kurtsiefer2001,Bovino2003,Bennink2010} or periodically poled crystals \cite{Ljunggren2005,Evans2011},
and in nonlinear waveguides \cite{Mosley2009}.

In the last two decades, a variety of HSP sources based on SPDC has been realized in an asynchronous configuration, via a CW pump
\cite{Fasel2004,Alibart2005,Castelletto2006,Tengner2007} and synchronously with a pulsed pump
\cite{Uren2004,Pittman2005,Castelletto2005,Soujaeff2007,Bussieres2008,Mosley2008}.
Recently, promising all-fibre synchronous HSP sources based on Four Wave Mixing (FWM) have also been reported \cite{McMillan2009,Slater2010,Soller2010,Soller2011}.
However, only some of these HSP sources are suited to quantum communication applications, where it is necessary to herald with high probability, and rate, the presence of photons at telecom wavelengths into a single mode fibre.
The majority of the HSP sources at telecom wavelengths based on SPDC operate in an asynchronous way \cite{Fasel2004,Alibart2005,Castelletto2006,Tengner2007}. To the best of our knowledge, the only HSP sources based on pulsed SPDC are reported in \cite{Soujaeff2007, Bussieres2008}, but suffer from poor heralding efficiencies.

Nonetheless, the pulsed pumping regime is more advantageous than the CW one in some quantum communication applications.
For example, in QKD the pulsed mode can make the synchronization of multiple communicating parties easier with respect to the CW mode.
Some other challenging multi-photon quantum information experiments can take advantage of the pulsed pumping configuration.
In particular, our study is motivated by the requirements of a recent proposal to use Sum Frequency Generation (SFG) between single photons
belonging to two distinct entangled photon pairs in order to faithfully herald entanglement at a distance \cite{Sangouard2011}. The pulsed pumping configuration, with respect to the CW one, allows one to synchronize the arrival of the two single photons in the nonlinear crystal performing the SFG.
For this experiment to be feasible, despite the low efficiency of the nonlinear interaction between single photons,
high repetition rates and efficient coupling are required.

However, the coupling efficiency of pulsed SPDC sources is usually limited by the group velocity mismatch between the
generated photons and the pump pulses inside the nonlinear crystal.
Indeed, because of the dispersive properties of the crystal, the SPDC generated photons can acquire a temporal delay with respect to the pump,
which can be larger than the pump pulses' coherence time. Under these conditions, the coherence of the photons emitted from
different longitudinal positions within the crystal is lost, resulting in a spatial multimode emission and in a reduction of the
fibre coupling. These effects can be limited by reducing the length of the crystal, which, in turn, causes a decrease of the source brightness.

In this paper, we choose the length of the crystal in order achieve a trade off between high brightness and high coupling efficiency.
Indeed, we experimentally demonstrate the realization of a synchronous HSP source at telecom wavelengths, based on pulsed SPDC in a 1\,cm
bulk PPLN crystal, with which telecom photons can be heralded at 4.4\,MHz rate with 45\% heralding efficiency.
To the best of our knowledge, this is the most efficient high rate HSP source based on pulsed SPDC ever reported.
We measure a 80\% efficiency of announcing a telecom photon, when we concentrate only on the coupling of the heralded photon.
Furthermore, by testing a 4\,cm crystal in a pulsed and CW pumping configuration, we experimentally investigate how the group velocity mismatch in longer crystals affect the spectrum and the coupling efficiency of the emitted photons.

\section{The source}\label{par:setup}

\begin{figure}[t]
\begin{center}
\includegraphics[width=1\textwidth]{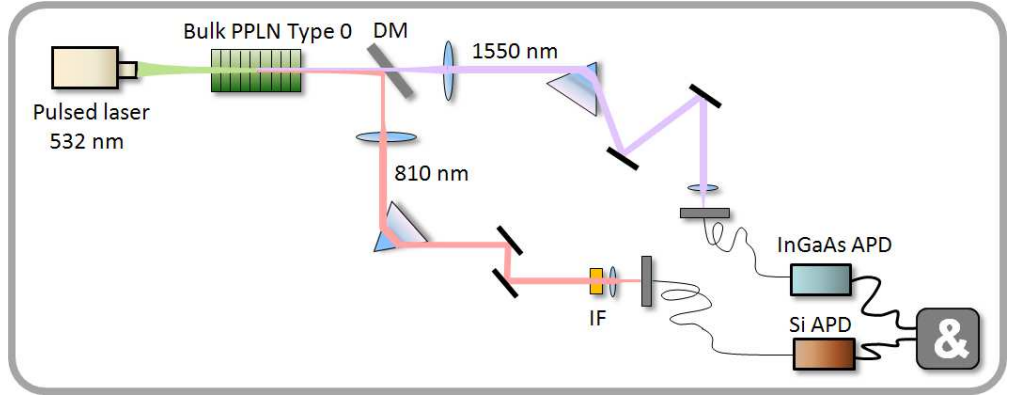}
\caption{\small Setup of the source. DM: dichroic mirror; IF: interference filter.}\label{fig:setup}
\end{center}
\end{figure}

A 430\,MHz passively mode-locked laser (Time-bandwidth GE-100-VAN-HP-SHG-430) producing pulses of 8\,ps at 532\,nm, pumps a 1\,cm
Type 0 (e,e,e) PPLN bulk crystal (Covesion), which is at a temperature of 180\,$^{\circ}$C (see figure \ref{fig:setup}), in order
to generate photon pairs at 810 and 1550\,nm via SPDC.
The pump laser is coupled into a polarization maintaining fibre in order to select a single spatial mode and focused on the center
of the crystal with an aspheric lens (Geltech 352110) with focal length f=6.24\,mm.

The generated signal (810\,nm) and idler (1550\,nm) photons are separated by a dichroic mirror and collimated using f=150\,mm
achromatic lenses (Thorlabs AC254-150-B/C). The pump light is removed using equilateral prisms (Schott F2 glass) at the Brewster angle.
The photons are then coupled into fibres using aspheric lenses (Geltech 352220-B for the idler with f=11\,mm and Geltech 350260-C for the
signal with f=15.36\,mm).
Injecting light backwards through the fibres allows one to check for the correct alignment, for the size of the waists in the crystal
and for the transmission in the optical path.
On the 810\,nm arm, we use an interference filter centered at 810\,nm, with a 10\,nm bandwidth and 55\% transmission,
to further remove the pump light.

The heralding photon at 810\,nm is detected by a free running Silicon Avalanche PhotoDiode (Si-APD) (Laser Components Count)
with $\eta_{810}=$50\% efficiency at 810\,nm and 5\,Hz of dark counts. The telecom photon is detected by an InGaAs APD (ID201, IDQuantique)
with $\eta_{1550}=$10\% efficiency and 8$\cdot 10^{-6}$ dark count probability per ns. Notice that we assume a 10\% relative uncertainty
in the calibration of the detectors efficiency, which produces a similar error in the coupling efficiencies reported in this paper.
In order to characterize the source, we measure via a Time to Digital
Converter (TDC) (Agilent Acqiris TC890) the coincidences between the detections of the InGaAs and the Si detectors as a function of their time
difference.

In order to study effects due to the group velocity mismatch in the presence of a pulsed pump, we also test a 4\,cm PPLN bulk crystal.
In this case, an aspheric lens (Geltech 350280-C) with f=18.40\,mm is used for fibre coupling the telecom photon and the distance
between the optical
elements is set in order to provide the optimal beam waists inside the longer crystal.
For further characterization of the source, a CW laser at 532\,nm (Laser Quantum Torus) is also used.

\subsection{Photon coupling optimization}\label{par:software_optimization}

The coupling efficiency of a single photon (or photon pair) source is a critical parameter, especially in multi-photon experiments where the probability of success decreases with the photon losses to the power of the number of photons involved in the experiment.
In \cite{Ljunggren2005} it is theoretically shown that by
properly focusing the pump onto a bulk crystal, one can attain SPDC photon emission very close to the
single mode of optical fibres, independently of the crystal length.
For a monochromatic pump with a specific waist, it is possible to calculate the density
matrix of the emitted SPDC two-photon state, which is constrained by phase-matching conditions, in terms of angular and frequency
distribution, as described in~\cite{Ljunggren2005}.
Due to the high dimensionality of the problem, the calculation can be numerically intensive but gives access to the full information about the state, including spatial and spectral correlations.
In our case, we use this information for optimizing the pump focusing in order to maximize the fibre coupling.

We calculate with this model that with a beam waist inside the 1\,cm crystal of 40\,$\mu$m for the pump and of 30\,$\mu$m for the idler and signal beams collected by the corresponding fibres, the photons generated by SPDC in the crystal are emitted with a spatial distribution which has a 96\% overlap with a single gaussian mode.
Beam waists of 80$\mu$m, 60$\mu$m, 60$\mu$m for the pump, signal and idler respectively are calculated for the 4 cm crystal.

Although there are optimal beam waists, the overlap between the pump and the generated photons can be better than 90\% over nearly one order of magnitude in waist size, provided that the collecting optics is chosen accordingly. In order to accurately plan the beam paths and to avoid optics or alignments which could introduce aberrations, we simulated the beams propagation using the optical modeling software "Code V". All alignment takes advantage of CCD cameras.

\section{Characterization of a HSP source}

Different parameters can be adopted for characterizing a HSP source. Here, we focus on the heralding efficiency and on the
conditional autocorrelation function.

\subsection{Heralding efficiency}\label{par:heralding_efficiency}
The heralding efficiency $\eta^{H}$ is the probability of finding the signal photon (1550\,nm) in the single mode fibre once the idler (810\,nm) is detected.
In our case, $\eta^{H}$ is equivalent to the overall transmission of the telecom photon, from the nonlinear crystal to the InGaAs detector.
A first and simple way to estimate $\eta^{H}$ is to measure the coincidences with the TDC between the InGaAs and the Si-APD detections, with the InGaAs detector triggered by the signal provided by the Si-APD. Indeed, in this situation the overall transmission of the telecom photon is given by:
\begin{equation}\label{eq:t1550}
t_{1550}=\eta^{H}=\frac{C}{S_{810} \eta_{1550}},
\end{equation}
where $C$ is the coincidence count rate and $S_{810}$ the singles on the Si-APD.
Accidental coincidences and dark counts of the Si-APD are subtracted from the value of $C$ and $S_{810}$ respectively.
It is desirable to perform the measurement of $t_{1550}$ at a low pumping regime, in order to make the emission of multiple photon pairs negligible.

The transmission of the photon at 810\,nm, can also provide other useful information about the source. It can be estimated by measuring the coincidences with the InGaAs APD in gated mode but not triggered by the Si detector.
Indeed, in this situation the transmission of the idler photon is given by
\begin{equation}
t_{810}=\frac{C'}{S_{1550} \eta_{810}}.
\end{equation}
where $C'$ are the net coincidence counts in this configuration and $S_{1550}$ the singles on the InGaAs-APD.

Once the overall transmission of the two photons is determined, together with the knowledge of the detectors' efficiency,
one can calculate the number of photon pairs generated per pulse $p$.
This quantity can be also measured by adopting an alternative method \cite{Marcikic2002}: the detection of the twin SPDC
photons produces the appearance of a coincidence peak in the distribution of the coincidences measured by the TDC with respect
to the time delay between the idler and the signal photon. However, for a pulsed source, because of inefficient detection,
side peaks, corresponding to coincidences between not correlated photons, are also present. The value of $p$ can also be determined
by measuring the ratio between the counts in one side peak and the counts in the main peak.


\subsection{Conditional autocorrelation function}
The conditional autocorrelation function $g^{(2)}_{a|b}(0)$ of the state of the heralded photon $a$, given a detection of the heralding photon $b$,
is often adopted as a figure of merit for evaluating the single photon emission of a HSP source. In our case, the photons $a$ and $b$ are
the idler at 1550\,nm and the signal at 810\,nm respectively.
For a SPDC source, the $g^{(2)}_{a|b}(0)$ can be calculated exactly using equation (24) in \cite{Sekatski2012}, that can be written in the case of low values of $p$ ($p<<1$) and no detector noise as
\begin{equation}\label{eq:g2}
g^{(2)}_{a|b}(0)=f_N(2-\eta_{b})p + O(p^2),
\end{equation}
where $f_N=1+\frac{1}{N}$ is a factor depending on $N$, i.e. the number of modes characterizing the state of the heralded photon, and $\eta_{b}$ is the overall detection efficiency of the triggering (heralding) photon $b$, which in our case can be expressed as
$\eta_{b}=t_{810} \eta_{810}$. The factor $f_N$ has a value of 2 in the single mode case ($N=1$), while it is equal to $1$ in the fully multimode configuration.
Notice that $f_N$ drops quickly to $1$ for increasing $N$.

The $g^{(2)}_{a|b}(0)$ also depends, to a first approximation, on the detection and coupling efficiency of the heralding
photon $\eta_{b}$. In particular, $g^{(2)}_{a|b}(0)$ increases for lower values of $\eta_{b}$. In \cite{Sekatski2012} the authors
take into account a non unit efficiency and non photon number resolving detector.

From equation (\ref{eq:g2}), we see that $g^{(2)}_{a|b}(0)$ is proportional,
to a first approximation, to the pair generation probability of the source $p$. Since $p$ represents the main factor on which
$g^{(2)}_{a|b}(0)$ depends, it is sufficient to measure $p$ in order to have a good estimation of the autocorrelation function $g^{(2)}_{a|b}(0)$.
The autocorrelation function $g^{(2)}_{a|b}(0)$ goes to zero for low values of $p$.
However, for practical applications, a tradeoff between a low $g^{(2)}_{a|b} (0)$ and a high photon generation rate must be found.
A pulsed laser with high repetition rate, such as the one used in our experiment, allows one to have high production rates even
at relatively low $p$ \cite{Broome2011}.

\section{Experimental results}

\subsection{High heralding rate}
We first show that with our source one can herald single telecom photons efficiently even at high heralding rates.
High heralding rates are allowed by the high repetition rate (430\,MHz) of our pump laser.
Moreover, they can be achieved with high detection and fibre coupling efficiency of the heralding photon at 810\,nm.
Therefore, we optimize the fibre coupling of both photons by maximizing the coincidence rate.

We initially characterize the source at low pump power, in order to avoid the saturation of the detectors and to measure more
accurately the overall transmission of the photons. At an average pump power of 45\,$\mu$W, we measure 49\,kHz of counts of the Si-APD
and 2.2\,kHz of coincidences with the InGaAs detector triggered by the Si-APD. Using equation (\ref{eq:t1550}),
we calculate a overall coupling, or heralding efficiency, of the telecom photon of 45\%.
We measure a 80\% transmission through the optical elements on the telecom arm of the source.
On the 810\,nm arm, we measure instead a overall coupling of 39\%.
Notice that the lower coupling efficiency at 810\,nm is partly due to the presence of additional filtering losses in the 810\,nm arm,
and to the fact that the collecting single mode fibre, unlike the 1550\,nm one, is not anti-reflection coated.
With the values mentioned above, we estimate to have generated a number of photon pairs per pulse of $p=6\cdot10^{-4}$. This value
is confirmed by the alternative measurement method described in \ref{par:heralding_efficiency}.
At this point, we increase the average pump power to 7.5\,mW, corresponding to $p=0.1$, and we measure 4.4\,MHz rate on the Si-APD.
Notice that this value is lower than what one might expect because of saturation effects of the Si-APD.

By using the equation (\ref{eq:g2}) in the multimode case, we calculate in this regime for the conditional autocorrelation function
$g^{(2)}_{a|b} (0)$ a
 value of $0.18$. This confirms that with our source it is possible to herald single telecom photons even at high heralding rates.

\subsection{High heralding efficiency} \label{par:source_characterization}
We now show that very high heralding efficiencies of the telecom photon can also be achieved with our source.
In this case, we optimize the coincidence rate by giving priority to the coupling of the telecom photon.

At an average pump power of 68\,$\mu$W, we measure 94\,kHz of counts of the Si-APD and 7.5\,kHz of coincidences with the InGaAs
detector triggered by the Si-APD. Using equation (\ref{eq:t1550}), we calculate a 80\% heralding efficiency of the telecom photon.
The photon at 810\,nm has an overall transmission of 5\%.
With the values mentioned above, we calculate $p=0.009$.

By using equation (\ref{eq:g2}), we find a value for the autocorrelation function $g^{(2)}_{a|b} (0)$ of $0.018$.
In this regime, corresponding to low values of $p$ and $100$\,kHz heralding rate, the heralded state corresponds of a single telecom
photon of good quality, in
the sense that it has a negligible multi-photon component.

\subsection{Group velocity mismatch in longer crystals} \label{par:spectrum}
In order to investigate the coupling efficiency with longer crystals, we test a 4\,cm PPLN crystal.
With an average pump power of 54\,$\mu$W, we measure 54\,kHz of counts of the Si-APD and 2.6\,kHz of coincidences with the InGaAs
detector triggered by the Si-APD.
The overall transmission of the telecom photon is 48\%.
We measure an overall transmission of the photon at 800\,nm of 24.5\%.
With these values we calculate a number of photon pairs per pulse $p$ of 0.001.
The difference in coupling efficiency between the 1\,cm and 4\,cm crystal can be attributed in part to the
difficulty of aligning the 4\,cm crystal which has a 0.5\,mm $\times$ 0.5\,mm clear aperture and in part,
as we shall see below, to the fact that the group velocity mismatch between the pump and the photons is more
critical in the longer crystal: the photons are not emitted coherently over the all length of the crystal,
and therefore they are not in a single spatial mode.

\begin{figure}[t]
\begin{center}
\includegraphics[width=1.0\textwidth]{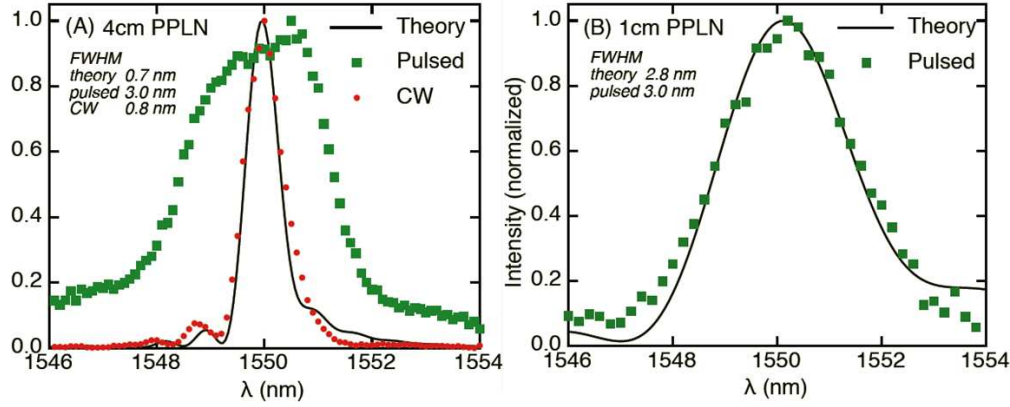}
\caption{Left: measured spectrum of the heralded photons in the pulsed and CW pumping configuration for a PPLN bulk crystal of 4\,cm of length. The calculated spectrum is also reported. Right: measured spectrum for a PPLN bulk crystal of 1\,cm of length in the pulsed regime compared with the theoretical curve.}\label{fig:spectrum}
\end{center}
\end{figure}

To illustrate the effects of group velocity mismatch in the long crystal, we characterize it in CW pumping mode, where these effects
are negligible. 
In this case we measure the overall transmission to be 60\% for the telecom photon and 20\% for the 810\,nm one. The 4\,cm source in a CW mode
shows a better fibre coupling with respect to when it is pumped in a pulsed mode. We measure the spectrum of the telecom
photons in the two pumping regimes (left part of figure \ref{fig:spectrum}).
The spectrum calculated by our software is in good agreement with that of the CW case. In the pulsed configuration, the spectrum is broader
and has a different shape. The model in \cite{Ljunggren2005} only considers a monochromatic pump, which is the case in the CW configuration,
therefore it does not
take into account dispersion effects for the pump pulses inside the crystal.
The difference between the pulsed and CW regime is related to the group velocity mismatch between the signal and the idler fields with respect
to the pump. The generated photons acquire, in the 4\,cm crystal, a temporal delay with respect to the pump that is larger than
the coherence time of the pump pulses. This causes a reduction of the effective crystal length in which the nonlinear interaction takes place.
Moreover, the single mode photon emission is deteriorated, causing a drop in the coupling efficiency.

Let us give a simple, yet illustrative estimation of the effect of the group velocity mismatch in the pulsed pumping mode.
The group velocity mismatch per unit length [s/m] is given by the difference of temporal delays of the telecom photon and of the pump pulses
\begin{equation}
\tau=\frac{1}{v_g^{532}}-\frac{1}{v_g^{1550}}=\frac{n_g^{532}}{c}-\frac{n_g^{1550}}{c},
\end{equation}
where $v_g^{532(1550)}$ and $n_g^{532(1550)}$ are the group velocities and group indices of the fields at 532 and 1550\,nm, and $c$
is the speed of the light in the vacuum.
By considering the PPLN dispersion properties \cite{Jundt1997}, we found, with $n_g^{532}=2.4746$ and $n_g^{1550}=2.1762$ at 180\,$^{\circ}$C,
a $\tau_{1550}=9.994$\,ps per cm. In a similar way, with $n_g^{810}=2.2697$, we find a delay of $\tau_{810}=6.830$\,ps per cm for the photon
at 810\,nm, with a mean delay of 8.412\,ps per cm.
The group velocity mismatch reduces the effective interaction length between the pump and the down-converted components to a value at which
the group delay is close to the coherence time of the pump pulses, which is around 8\,ps. Therefore, the group velocity mismatch turns out to be
less critical in a PPLN crystal of 1\,cm pumped by our laser.
Indeed, in the right part of figure \ref{fig:spectrum}, we report the spectrum of the telecom photons
produced by the 1\,cm crystal and with the pulsed pump. In this case, the spectrum is in better agreement
with the one calculated without considering the group velocity mismatch.

Taking into account these effects seems very important for optimizing the fibre coupling in a pulsed regime for SPDC sources.
For this reason an extension of the theoretical model in \cite{Ljunggren2005}, taking into consideration the spectral
and temporal properties of the pump should be formulated.

\section{Comparison with other HSP sources}

In Table \ref{tab:comparison_HSPS} the most relevant synchronous and asynchronous HSP sources
based on SPDC and FWM at telecom wavelength are compared. Different parameters are considered:
\begin{itemize}
\item the spectral bandwidth of the heralded telecom photon ($\Delta\lambda^{H}$);
\item the average pump power ($P_{pump}$ );
\item the number of photon pairs generated per pulse ($p$) in the case of synchronous sources, or, alternatively, the number of pairs
generated per ns for the asynchronous sources;
\item the reported heralding rate ($R^H$ and rate $R^H_{p=0.1}$ in the case of $p=0.1$);
\item the heralding efficiency ($\eta^{H}$);
\end{itemize}

\begin{table}[t]
\centering
\vspace{0.5cm}
\begin{tabular}{c|l|cccccccc}
\hline
\hline
&& Year    & Process        &   $\Delta \lambda^{H}$          & $P_{pump}$                & p       & $R^H$  & $R^H_{p=0.1}$  &   $\eta^{H}$ \\
&&         &                         &   [nm]                           &  [mW]           &         &  [kHz] &  [kHz]&         [\%] \\
\hline
&This work                 & 2012 &  SPDC   &   3&  7.5  & 0.1  &   4.4 $\cdot 10^{3}$  & 4.4 $\cdot 10^{3}$ $^{*}$ &    45$\pm 5$\\
&  This work                  & 2012 &  SPDC    &  3  &  0.068              & 0.009  &   94   &  &  80$\pm8$\\
\begin{sideways}C\end{sideways}
&S\"oller \cite{Soller2010}         & 2010 &  FWM      &   35        &    0.1             & 0.08     &   16.5    &   26   &    28       \\
\begin{sideways}N\end{sideways}&Slater \cite{Slater2010}           & 2010 &  FWM      &   5.8       &    20              & NR     &   54   &   -   &    25       \\
\begin{sideways}Y\end{sideways}&McMillan \cite{McMillan2009}       & 2009 &  FWM      &   0.8       &    65              & 0.094  &   92   &97&    52       \\
\begin{sideways}S\end{sideways}&Bussi\`eres \cite{Bussieres2008}   & 2008 &  SPDC     &   15        &    0.5             & 0.04  &   25   & 63 &    12       \\
&Soujaeff \cite{Soujaeff2007}       & 2007 &  SPDC     &   18        &    240             & 0.083  &   216  & 260 &    19     \\
\hline
\begin{sideways}C\end{sideways}& & &    &              &            &     &   & &     \\
\begin{sideways}N\end{sideways}&Tengner \cite{Tengner2007}         & 2007 &  SPDC    &  7                &    1.2             &   1.3 $\cdot 10^{-3}$/ns      &  81   & 6 $\cdot 10^{3}$ &    48  \\
\begin{sideways}Y\end{sideways}&Castelletto \cite{Castelletto2006} & 2006 &  SPDC    &   4                 &    100              &    NR   &  NR   & - &    48    \\
\begin{sideways}S\end{sideways}&Alibart \cite{Alibart2005}         & 2005 &  SPDC   &   20                 &    $10^{-2}$       &   NR     &  104 & -  &   37    \\
\begin{sideways}A\end{sideways}&Fasel \cite{Fasel2004}             & 2004 &  SPDC   &  6.9                &    49              &    $\approx 1.3 \cdot 10^{-2}$/ns$^{**}$   & 845   & $\approx 6.5 \cdot 10^{3}$&   60  \\

\hline
\end{tabular}
\caption{\small Comparison between HSP sources based on SPDC and FWM at telecom wavelengths.
Synchronous (SYNC) and asynchronous (ASYNC) HSP sources are listed in the upper and lower
block of the table respectively.  'NR' stands for not reported. $^{*}$This is a measured value,
while the the others reported in same column are calculated. $^{**}$A SPDC conversion efficiency of $10^{-10}$
is considered in this case. See text for other details.  }
\label{tab:comparison_HSPS}
\end{table}

The results in Table \ref{tab:comparison_HSPS} suggest that HSP sources based on SPDC have better heralding
efficiencies when working in an asynchronous way. All asynchronous sources listed in Table \ref{tab:comparison_HSPS}
are based on periodically poled crystals, except for the source in \cite{Alibart2005}, made of a nonlinear waveguide.
However, this source shows only a 37\% coupling, confirming the fact that the waveguide technology still needs optimization
in terms of internal losses and single mode operation.

Among the synchronous HSP sources at telecom wavelengths reported in the literature, only two, as far as we know, are based on SPDC
 \cite{Soujaeff2007,Bussieres2008}. However, these sources show poor heralding efficiencies of 19\% and 12\% respectively.
In the last few years, more efficient synchronous HSP sources have been obtained with fibre-based FWM, such as in \cite{McMillan2009}.
However, the source described in this paper, producing single telecom photons at MHz heralding rates with 45\% heralding efficiency,
turns out to be the fastest and the most efficient among the synchronous HSP sources in the telecom regime.

\section{Conclusion}
In this paper we have described a synchronous SPDC source of heralded single photons at telecom wavelengths with high heralding efficiency
and MHz heralding rates.
This source, which uses a 1\,cm PPLN crystal pumped at 532\,nm and generating photons at 810\,nm and 1550\,nm, achieves a 45\%
heralding efficiency with a 4.4\,MHz heralding rate, and 80\% heralding efficiency when we focus only on the coupling of the heralded photon.
These are the best results reported so far for a source of heralded telecom photons based on SPDC and a pulsed pumping regime.
Moreover, they show that even with a pulsed pump one can obtain high fibre couplings. The length of the nonlinear crystal
has to be chosen in order to make a trade off between high coupling, limiting the issues related to the group velocity mismatch for a given
pulsed pumping configuration, and high brightness.
Our synchronous HSP source could be well suited to building up multi-photon quantum communication experiments in which low-losses and
high count rates are demanded.

\section*{Acknowledgments}
We would like to thank valuable discussions with Felix Bussi\`eres, Nicolas Sangouard, Pavel Sekatski and Nicolas Gisin.
Financial support for this project has been provided by the Swiss National Science Foundation (SFNS) and by Q-Essence.

\end{document}